# Slip length in shear flow over a textured surface

Nicolas Elie[1]*, Pascal Jolly[1], Romain Lucas-Roper[2], and Noël Brunetière[1]

1) Institut Pprime, CNRS-Université de Poitiers-ISAE ENSMA

1-1 SP2MI, Bât. H1, 11 Boulevard Marie et Pierre Curie, TSA 41123, 86073 Poitiers Cedex

2) University Limoges, IRCER, UMR CNRS 7315, 87000 Limoges, France

2-2 Centre Européen de la Céramique, 12 Rue Atlantis, 87068 Limoges, France

*Corresponding author: noel.brunetiere@univ-poitiers.fr

Abstract

Hydrophobic textured surfaces are studied for their low wettability and their capacity to create a 'slippery' fluid on the surface during lubrication. To this end, the flow between two parallel surfaces is numerically addressed by computing two dimensional numerical simulations. One of the surfaces moves with a uniform rectilinear motion, while the other is fixed, with a cavity in the middle. The steady-state flow is laminar and monophasic with a low Reynolds number. The reduction of the wall shear stress caused by a vortex in the cavity, with respect to a Couette flow, looks like the creation of an equivalent slip of the fluid on the wall at a macroscopic scale. Three methods are used to calculate the slip length: one is based on the wall shear stress and the other two are based on the speed of the fluid flow. When the slip length is calculated according to these three methods, the obtained results differ. The differences show that the slip often used in the literature is a macroscopic representation of local effects that are not necessarily slippery. The speed profiles and the streamlines are then discussed, in order to propose an explanation for this difference.

**Keywords: slip length, drag reduction, textured surfaces, shear flow, parametric study**

# 1 Introduction

The reduction of friction and wear in lubricated conditions is a major challenge in the current context, where energy saving is a necessity [1]. Surface texturing is known for its influence on the potential reduction of the drag force between surfaces under shear flow on lubricated contacts [2]. Two parameters are important when studying such a configuration: the lift force influencing the height of the fluid film and the drag force. For example, in a theoretical and experimental analysis, Adjemout et al. [3] obtained a 50% reduction of friction with textures on the surface of a mechanical seal. Surface texture permits the generation of a more important lift force than with a non-textured mechanical seal, allowing for full-film lubrication between the seal surfaces. There are many papers concerning surface texturing and different kinds of textures make the optimization process of texture geometry difficult [4]. Since the ways of manufacturing texture have improved over time, new forms and sizes of texture could be possible in the future [5]. This means that many possibilities are not well known, or have never been studied, either in a numerical or an experimental way.

Another mechanism of friction reduction, based on lubricant slipping on the surfaces, was proposed by Spikes in 2003 [6,7]. From numerical simulations, Salant and Fortier [8] demonstrated that a lift force can be generated on a slider bearing, by having a successive slip zone and non-slip zone on a surface. The level of fluid slip on a surface is quantified by the slip length, which is defined as the distance between the wall and the point where the speed of the fluid would become zero, by the extrapolation of the speed profile when the fluid does not adhere to the wall. The slip length is correlated to the surface texture and the wetting of the surface, with higher slip lengths for super-hydrophobic surfaces [9]. The wetting is measured by the contact angle made by a sessile drop on the surface, a contact angle higher than 90° corresponding to hydrophobic surfaces [10]. Very high contact angles (super-hydrophobic state) are reached when the surface is textured and has a low surface energy. In this case, the liquid can fully (the Wenzel state) or partially wet the textured surface (air in the texture, the Cassie-Baxter state) [10].

Choi and Kim [11] obtained a slip length of about 50 μm with a super-hydrophobic textured surface coated with PTFE (PolyTetraFluoroEthylen) Fl. The slip length was measured with a cone-and-plate rheometer, on the basis of the friction reduction compared to a theoretical friction without slip. They analyzed both hydrophobic and hydrophilic surfaces in both a smooth and a textured configuration. The slip length was only obtained for the case with a textured hydrophobic surface. In this article, the fluid was water with 30% glycerin. Lee et al. [12] published a review article about the drag reduction in a laminar flow in hydrophobic conditions. One of their conclusions was that several important issues have to be addressed before deploying such surfaces, including the resistance of the Cassie-Baxter state to higher pressures. Solomon et al. [13] carried out a study with lubricant-impregnated textured surfaces. These surfaces were used in shear flow with a second immiscible liquid, to maintain the first fluid in the texture. The obtained drag reduction was 16%, corresponding to a slip length of 18 μm, when the impregnated lubricant had a viscosity more than 200 times lower than the working fluid. Jendoubi et al. (2018) [14] used a super-hydrophobic textured PTFE surface in a sliding contact but obtained no drag reduction. According to the authors, the textures were too wide to maintain the air in the textured cavities. They suggested the use of very narrow cavities (of sub-micrometric size).

The size of the textures could be reduced by new manufacturing techniques, including self-assembly [15]. The problem is that the experimental observation of the behavior of the fluid is not easy (even impossible) to do at a submicroscopic scale. As an example, the particle image velocimetry technique used to obtain the streamlines is limited by the size of the particles. Considering these limits, simulation could be a solution to understanding the behavior of the fluid at these scales.

When considering depth to width ratios of cavities greater than 1, or textures that help to reach the super-hydrophobic state and slip [16], the Reynolds equation cannot be used for this type of problem; computational fluid dynamics (CFD) must be considered [17].

Arghir et al. [18] conducted CFD simulations of textured surfaces in shear flows and found that a load carrying capacity can be achieved. Sahlin et al. [19] performed a numerical analysis of the influence of a cavity with a cylindrical and a splined geometry on drag and lift force. They concluded that the maximal lift capacity is reached close to the moment where an eddy is created inside the cavity. The drag force decreases when the depth and the diameter of the cavity are increased. Dobrica and Fillon [17] explained that the observed load carrying capacity was misleading because of the way the reference pressure was chosen.

Caramia et al. [20] carried out a numerical study on a two-dimensional cavity by increasing its depth. It was shown that the depth influences the normalized wall shear stress by the unit length, the dynamic viscosity and the speed, for three heights of the film. The results were compared to the experimental findings from Scaraggi et al. [21]. A reduction in drag force, of about 20%, was obtained by simulation, whereas it was 80% in the experimental study. The authors considered that the discrepancy was due to cavitation or air entrapment in the cavities.

In order to have a better understanding of the behavior of such systems, we carried out a parametric study of a single cavity in a two-dimensional configuration and in the Wenzel wetting state (i.e., when the texture is fully wetted). The obtained drag force was compared to that for a Couette flow between parallel surfaces. The apparent slip length was estimated for different geometrical configurations of the cavity using different methods. It is shown that the calculation of slip length from the shear stress, as carried out in many papers, is not correct.

## 2 Methods

2.1 Description

The configuration of the two-dimensional problem studied in this paper is presented in Fig. 1. A moving wall with a speed $U_x$ was placed at a distance $h_f$ over a textured surface. Non-dimensional parameters were introduced to describe the problem:

- Aspect ratio of the cavity: $w = \frac{h_c}{d}$, where $h_c$ and $d$ are the depth and the width of the cavity, respectively.
- Size ratio of the cavity: $\frac{h_f}{d}$.
- Density of textures: $\frac{d}{l_s}$, where $l_s$ is the length of the domain.
- Reynolds number: $R_e = \frac{U_x h_f}{\nu}$.

$\nu$ is the kinematic viscosity of the fluid in the cavity.

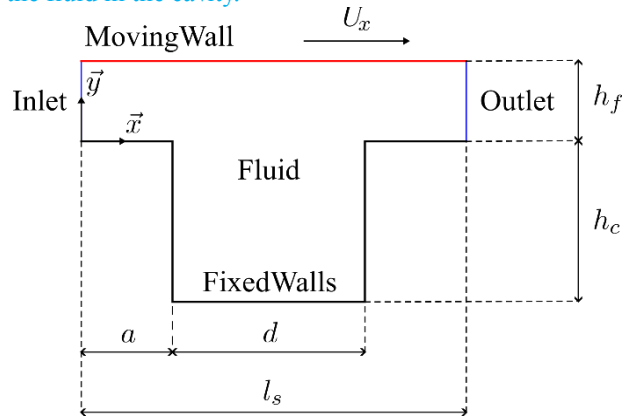

**Fig. 1 Configuration of the two-dimensional cavity for the numerical analysis.**

Navier-Stokes equations were solved using the finite volume method with the simpleFoam solver from OpenFOAM 9 [22]. The solver uses the SIMPLEC (Semi-Implicit Method for Pressure Linked Equations-Consistent) algorithm and is an iterative method based on the well-known SIMPLE (Semi-Implicit Method for Pressure Linked Equations) algorithm [23], where the velocity correction procedure is modified compared to the SIMPLE method. The system works without cavitation and the surfaces are considered to be perfectly wetted by the liquid. The moving wall has a speed $U_x$ on the $\vec{x}$ axis, corresponding to a Reynolds number of 0.733. A periodic condition was applied between the outlet and the inlet; the walls had a no-slip condition. In this steady-state problem, the fluid was considered to be incompressible and Newtonian, with water properties at a temperature of 20°C. The texturing density was kept constant at 0.5 and the aspect ratio varied from values of $w \ll 1$ to $w = 4$; the size ratio varied from $\frac{h_f}{d} = 0.02$ to $\frac{h_f}{d} = 2.00$. In this first study, cavitation was neglected because it only occurs when the average fluid pressure is close to atmospheric pressure. As the application targeted by this study was seals, high levels of fluid pressure were generally encountered, eliminating the risk of the occurrence of fluid cavitation.

2.2 Reference pressure

As the system is bounded by a periodic condition, the pressure is only relative and needs a reference in the CFD software. In the present case, the pressure was kept at zero at the center of the coordinate system, as shown in Fig. 1.

2.3 Slip length

The speed profile of the fluid inside the cavity is presented in Fig. 2 a. Inside the cavity, the speed profile corresponds to a vortex. The average of the speed profile along the $\vec{x}$ axis gives the speed profile in Fig. 2 b, where the slip length $\delta$ is defined as the extrapolation of the speed profile in the range $y \geq 0$, to the point where it becomes zero.

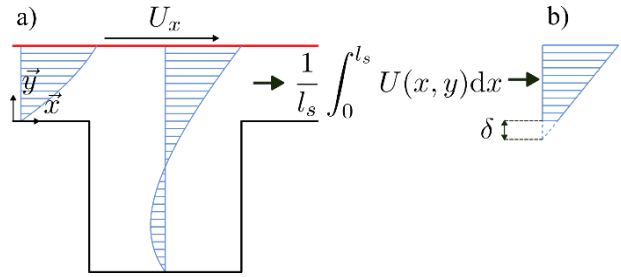

**Fig. 2 a) Speed profile of the fluid and b) slip length definition**

The apparent slip of the flow reduced the shear stress and increased the average speed of the flow, compared to the case of a shear flow for which the shear stress was $\tau = \frac{\mu U_x}{h_f}$ and the average speed was $\frac{U_x}{2}$. The slip length was calculated by three different methods, which are illustrated in Table 1, and based on the average wall shear stress and average speed. The average wall shear stress is given by Eq. (1) ($\tau$ being the wall shear stress). The dimensional slip length was calculated from the average wall shear stress by Eq. (4) ($\mu$ being the dynamic viscosity) and the non-dimensional slip length (named 'slip proportion') was obtained when it was normalized by $h_f$ (Eq. (7)). Two average speeds of fluid flow were considered: the first was the average inlet speed (Eq. (2)) and the second was the average speed profile for $y \geq 0$ (Eq. (3)), where $U$ is the speed of the flow in the $x$ direction. The slip length was calculated from average inlet speed using Eq. (5) and the average speed profile was calculated from Eq. (6). The slip proportion obtained from the average inlet speed and average speed profile are given by Eq. (8) and Eq. (9), respectively.

**Table 1 Comparison of the slip length calculation methods**

|  | From wall shear stress | From average inlet speed | From average speed profile |
|---|---|---|---|
| Average wall shear stress calculation | $\tau_{av} = \dfrac{1}{l_s}\int_0^{l_s} \tau\, dx$ (1) | | |
| Average speed calculation | | $U_{xav} = \dfrac{1}{h_f}\int_0^{h_f} U(y)\, dy$ (2) | $U_{xp} = \dfrac{1}{h_f l_s}\int_0^{h_f}\int_0^{l_s} U(x,y)\, dx\, dy$ (3) |
| Dimensional slip length | $\delta_\tau = \dfrac{\mu U_x}{\tau_{av}} - h_f$ (4) | $\delta_U(U_{xav}) = h_f\left(\dfrac{\dfrac{U_{xav}}{U_x} - \dfrac{1}{2}}{1 - \dfrac{U_{xav}}{U_x}}\right)$ (5) | $\delta_U(U_{xp}) = h_f\left(\dfrac{\dfrac{U_{xp}}{U_x} - \dfrac{1}{2}}{1 - \dfrac{U_{xp}}{U_x}}\right)$ (6) |
| Non-dimensional slip length | $\bar{\delta}_\tau = \dfrac{\delta_\tau}{h_f}$ (7) | $\bar{\delta}_U(U_{xav}) = \dfrac{\delta_U(U_{xav})}{h_f}$ (8) | $\bar{\delta}_U(U_{xp}) = \dfrac{\delta_U(U_{xp})}{h_f}$ (9) |

2.4 Validation

Arghir et al. [18] used the SIMPLE algorithm to solve the Navier-Stokes equations numerically. To validate the present model, a comparison was made with their results. The parameters used in the simulation are given in the caption of Fig. 3. Fig. 3 shows the non-dimensional pressure profiles obtained by the present model and that of the reference, as a function of the non-dimensional $\bar{x}$ coordinate. The two curves are very close, suggesting that the present model is sufficiently accurate.

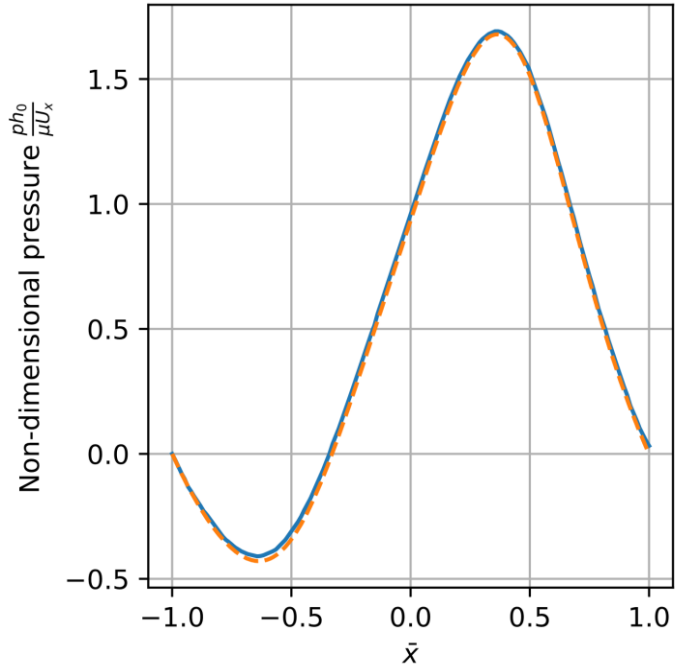

Fig. 3 Comparison of the non-dimensional pressure profile with the one obtained by Arghir et al. [18] versus the non-dimensional $\bar{x} = \frac{2x}{l_s}$ axis. The non-dimensional parameters are $\frac{d}{l_s} = 0.5$, $\frac{h_c}{h_f} = 2.0$ and $Re_{h_0} = \frac{\rho U_x h_0}{\mu} = 100$, where $h_0 = h_f + \frac{h_c}{2}$. The continuous line represents the reference and the dotted line represents the comparison with the present model.

3 Results

3.1 Wall shear stress profile

The dimensionless wall shear stress profiles on the moving wall for different size ratios and a constant aspect ratio $w = 2$ are given in Fig. 4. For the lowest size ratio (0.02), the wall shear stress is constant in the non-textured part, i.e., in the range [-1.0, -0.5] and [0.5, 1.0]. Above the cavity, the wall shear stress decreases down to values 10 times lower than those without any cavity. At the edges of the cavity, local increases are observed. When the width of the cavity decreases, in comparison to the film thickness, the shear stress spikes spread across the non-textured area and above the cavity. When the width of the cavity is close to the film height (size ratio = 0.5 and 2.0), the two-localized increase at the edges of the cavity merge and this leads to a maximum wall shear stress in the center of the cavity. Generally speaking, the wall shear stress is lower than the value in the non-textured case (i.e., lower than 1). Only the shear stress spike due to the cavity edges leads to a local shear stress higher than 1.

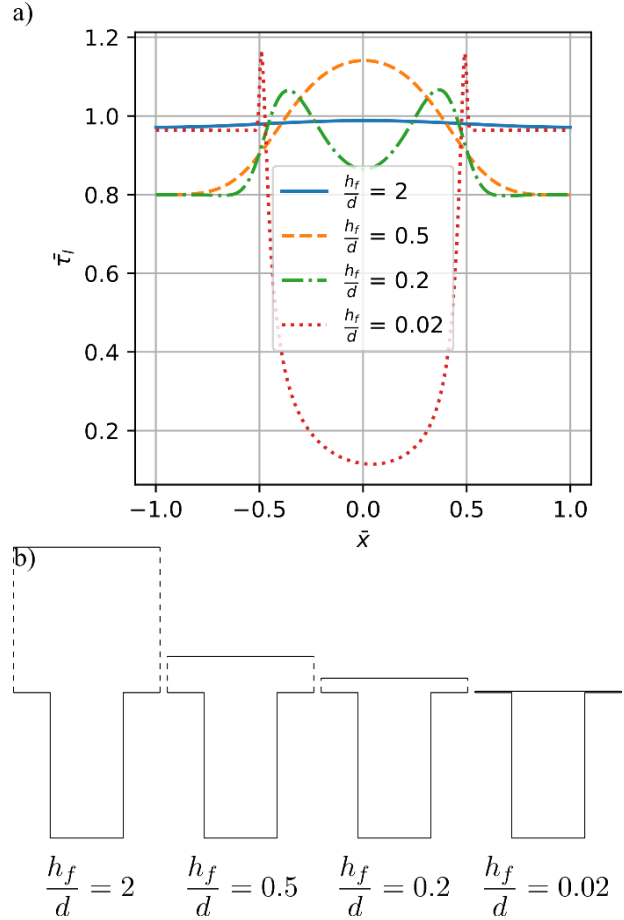

**Fig. 4** Non-dimensional local wall shear stress $\bar{\tau}_l = \frac{\tau_l h_f}{\mu U_x}$ on the moving wall for different size ratios of the cavity, as a function of the non-dimensional coordinate $\bar{x} = \frac{2x}{l_s}$ with a constant aspect ratio $w = 2$

3.2 Average wall shear stress

As the wall shear stress profile is not uniform, it is important to see how the average value (which is linked to the apparent slip) varies with the texture dimension. On Fig. 5, the wall shear stress is averaged along the moving wall and normalized by the value of the Couette flow with the same film height $h_f$ ($\frac{\mu U_x}{h_f}$); it is plotted as a function of the cavity aspect ratio for different size ratios. The size ratio varies from 0.02, when the width of the cavity is large compared to the height of the film, to 2.00, when it is half of the film height. Generally speaking, the dimensionless wall shear stress with a cavity is less than 1, i.e., the shear stress value without a cavity. The wall shear stress decreases when the depth of the cavity and the width of the cavity increase. As in the paper by Caramia et al. [20], two zones where the wall shear stress does not depend on the height of the cavity are observed. The second shear stress plateau corresponds to asymptotic behavior, occurring when the height of the cavity is equal to or higher than the width ($\frac{h_c}{d} \geq 1$). The first plateau appears at lower (but varying) values of the aspect ratio. The data are plotted as a function of another parameter, $\frac{h_c}{h_f}$, named the 'depth ratio' in Fig. 6. This parameter suggests that, when the height of the cavity is sufficiently small compared to the film height, the drag reduction almost does not depend on the width of the cavity. More interestingly, when the size ratio is sufficiently small ($\frac{h_f}{d} \leq 0.2$), the first wall shear stress plateau is close to 0.92 and occurs at depth ratio values of about 1.00.

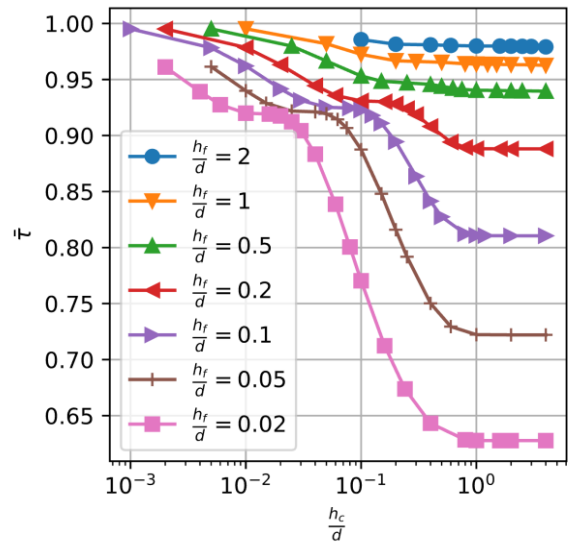

**Fig. 5 Non-dimensional averaged wall shear stress on the moving wall $\bar{\tau}$ as a function of the aspect ratio $\frac{h_c}{d}$ for different values of the size ratio $\frac{h_f}{d}$.**

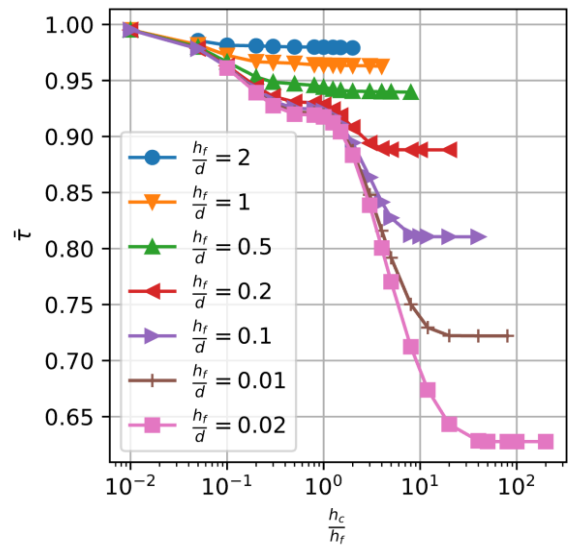

**Fig. 6 Non-dimensional wall shear stress $\bar{\tau}$ as a function of the depth ratio $\frac{h_c}{h_f}$ for different values of the**

size ratio $\frac{h_f}{d}$.

Fig. 7 a to c show the streamlines for a constant size ratio $\frac{h_f}{d} = 0.2$ and three different aspect ratios. In Fig. 7 a, corresponding to a low aspect ratio, two small vortices exist at the bottom edges of the cavity. The streamlines from inlet to outlet are entering the cavity and interacting with the bottom wall. When the aspect ratio increases (Fig. 7 b), the two vortices merge in one vortex at the bottom of the cavity. The streamlines from inlet to outlet are still entering the cavity but without interacting with the bottom wall due to the vortex. This corresponds to the first plateau, where $\frac{h_c}{h_f} = 0.8$ is close to one. When the aspect ratio keeps increasing, the size of the vortex increases until the height of the cavity is about the same as the width of the cavity. After this step, the center and the size of the vortex do not vary as well as the wall shear stress (Fig. 7 c); this corresponds to the second plateau. In this situation, the vortex is large enough to have its top streamlines going above the top of the cavity. Fig. 7 d to f show the streamlines for the size ratio $\frac{h_f}{d} = 0.5$ and different aspect ratios. It would seem that increasing the aspect ratio leads to the formation of a vortex (Fig. 7 a to c) but, this time, the top of the vortex never goes beyond the top of the cavity.

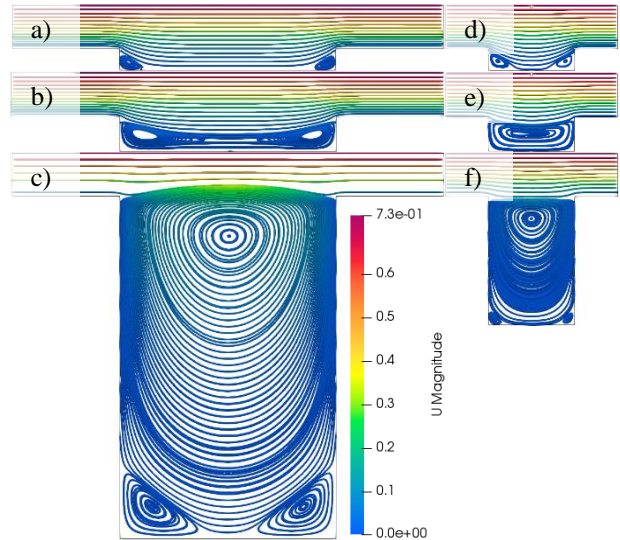

**Fig. 7 Streamlines and speed (m/s) in a cavity for a constant size ratio $\frac{h_f}{d} = 0.20$ where a) $w = \frac{h_c}{d} = 0.10$, b) $w = 0.16$, and c) $w = 1.60$ and a constant size ratio $\frac{h_f}{d} = 0.50$ with d) $w = 0.25$, e) $w = 0.40$ and f) $w = 1.50$.**

3.3 Slip proportion

As discussed in the previous section, the average wall shear stress for a textured surface is lower than for a flat surface. This will lead to an apparent slip. The slip proportion obtained with the wall shear stress and the two other methods, based on average speed, were analyzed. In Fig. 8, the slip proportions calculated by the three methods ($\bar{\delta}_\tau$, $\bar{\delta}_U(U_{xav})$ and $\bar{\delta}_U(U_{xp})$) are plotted versus the size ratio $\frac{h_f}{d}$ with a constant aspect ratio $w = 2$. The slip proportions calculated from the wall shear stress $\bar{\delta}_\tau$ and from $\bar{\delta}_U(U_{xp})$ are shown to be the same. They increase when the width of the cavity $d$ increases, compared to the film height $h_f$. Interestingly, the slip proportion calculated from the average inlet speed $\bar{\delta}_U(U_{xav})$ evolves in a different way to the other two slip proportions. When $\frac{h_f}{d} \geq 0.5$, the slip proportion calculated from the average inlet speed $\bar{\delta}_U(U_{xav})$ is slightly higher than the other slip proportion values, $\bar{\delta}_\tau$ and $\bar{\delta}_U(U_{xp})$. On the other hand, when the width of the cavity is sufficiently large, compared to the height of the film ($\frac{h_f}{d} \leq 0.2$), the slip proportion calculated from the average speed ($\bar{\delta}_U(U_{xav})$) tends to zero. This interesting difference is discussed in Section 4.

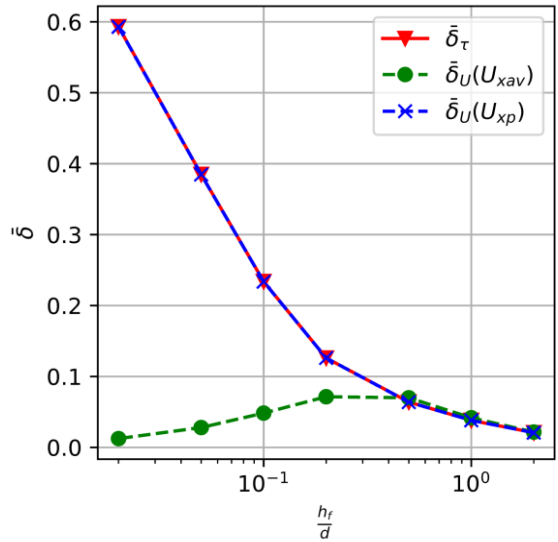

**Fig. 8 Comparison of the slip proportion calculated from wall shear stress $\bar{\delta}_\tau$, with that calculated from the average speed of the fluid $\bar{\delta}_U(U_{xav})$ and from the average speed profile in the lubricant film fluid $\bar{\delta}_U(U_{xp})$ versus the size ratio (with a constant aspect ratio $w = 2$.)**

3.4 Pressure profile

The texture and the resulting apparent slip will affect the pressure distribution along the moving wall, compared to a shear flow between parallel surfaces for which the pressure is uniform. The pressure variation can have a positive effect on the load generation between the surfaces, as discussed in the introduction. The non-dimensional pressure profile on the moving wall, for four size ratios and an aspect ratio $w = 2$, is presented in Fig. 9 as a function of the non-dimensional coordinate $\bar{x} = \frac{2x}{l_s}$. For each case, a pressure decrease is obtained at the leading edge of the cavity, with a pressure increase at the trailing edge. The shape and height of the pressure spikes change with the size ratio of the cavity, $\frac{h_f}{d}$. When the width of the cavity is sufficiently small, compared to the film height ($\frac{h_f}{d} \geq 0.2$), the profile is almost antisymmetric. For lower values of size ratio, a negative pressure is obtained in the center of the cavity.

A pressure gradient can be calculated from $\frac{p_{max}-p_{min}}{d}$, where $p_{max}$ and $p_{min}$ are the maximum and minimum pressure on the moving wall, respectively. Fig. 10 shows the dimensionless pressure gradient $\frac{p_{max}-p_{min}}{dP}$, with $P = \frac{6\mu U_x}{h_f^2}$ as a function of the depth ratio $\frac{h_c}{h_f}$ for different values of size ratio $\frac{h_f}{d}$. For each case, the pressure gradient increases from zero to a peak and then decreases to a plateau. The magnitude of the peak increases with the width of the cavity. The value of the pressure gradient at the plateau is at a maximum when the size ratio is between 0.2 and 0.5, corresponding to the maximum of the slip proportion $\bar{\delta}_U(U_{xav})$.

In Fig. 11, $\frac{p_{max}-p_{min}}{dP}$ is plotted versus the non-dimensional inlet speed of the fluid $\bar{U} = \frac{U_{xav}}{U_x}$, for different size ratios $\frac{h_f}{d}$ of the cavity. The curves for each ratio appear to be almost superposed and proportional to the average speed of the fluid, except for $\frac{h_f}{d} = 2$, where the slope looks to be the same but a constant downward shift is observed. From the Reynolds equation, it is possible to get Eq. (10), relating the pressure gradient to the average inlet speed:

$$\frac{dp}{dx} = \frac{6\mu U_x}{h_f^2}\left(2\frac{U_{xav}}{U_x} - 1\right) \quad (10)$$

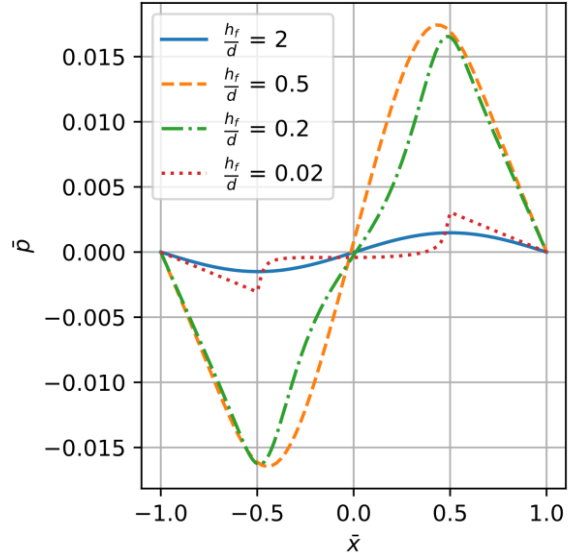

**Fig. 9 Dimensionless pressure $\bar{p} = p\frac{h_f^2}{6\mu U_x l_s}$ on the moving wall for different size ratios of the cavity versus the non-dimensional coordinate $\bar{x}$, with a constant aspect ratio $w = 2$.**

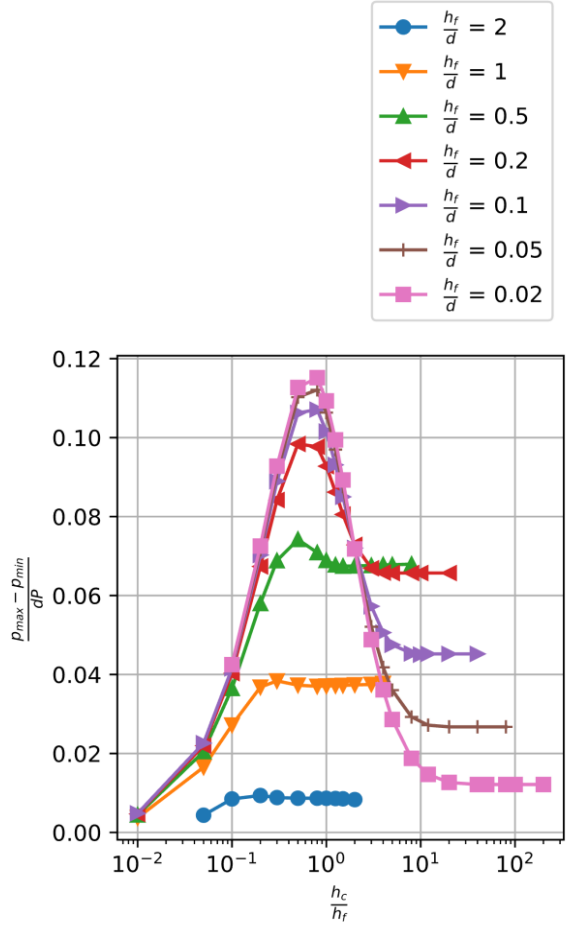

**Fig. 10 Dimensionless pressure gradient** $\frac{p_{max}-p_{min}}{dP}$ **with** $P = \frac{6\mu U_x}{h_f^2}$ **versus the depth ratio** $\frac{h_c}{h_f}$ **for different values of the size ratio** $\frac{h_f}{d}$**.**

Equation (10) is plotted in Fig. 11 and is very close to the curves obtained from the simulations. The case $\frac{h_f}{d} = 2$ corresponds to a configuration where the length of the domain $2d$ is of the same order of magnitude as its height. Thus, the Reynolds equation does not apply, explaining the difference observed with the other results.

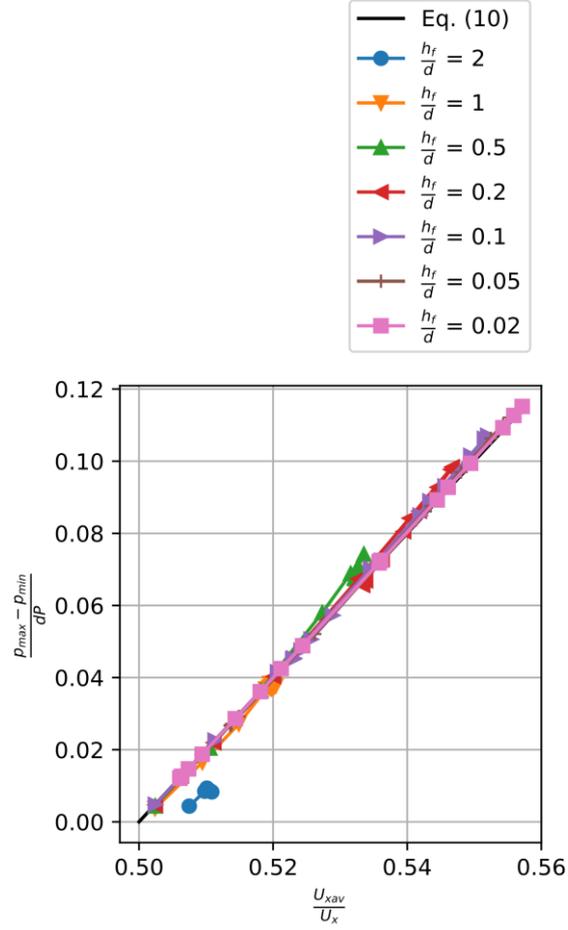

Fig. 11 Dimensionless pressure gradient $\frac{p_{max}-p_{min}}{dP}$ with $P = \frac{6\mu U_x}{h_f^2}$ versus the dimensionless inlet fluid speed for different size ratios of the cavity

**4 Discussion**

4.1 Slip proportion

The differences in the slip or, more accurately, the apparent slip proportion calculated with the three different methods and presented in Fig. 8, show the limits of the notion of slip length defined in the literature. The slip proportion calculated from the wall shear stress $\bar{\delta}_\tau$ increases when $\frac{h_f}{d}$ decreases but the slip length calculated from the average speed $\bar{\delta}_U(U_{xav})$ shows the same trends, down to $\frac{h_f}{d} = 0.5$, and then decreases. This is caused by the fact that the average speed is affected by the speed and the size of the vortex in the cavity (see Fig. 7). Fig. 12 a and b correspond to size ratios $\frac{h_f}{d} = 0.5$ and $\frac{h_f}{d} = 0.2$, respectively, and give an idea of the effect of the size ratio on the transition in $\bar{\delta}_U(U_{xav})$ behavior. The transition seems to happen when the top of the vortex is above the top of the cavity. When the top of the vortex is above the top of the cavity, $\frac{h_f}{d} = 0.2$, the slip proportion decreases when the size ratio is reduced. The speed of the vortex helps to maintain a high speed in the main channel. In the other case, the opposite behavior is observed.

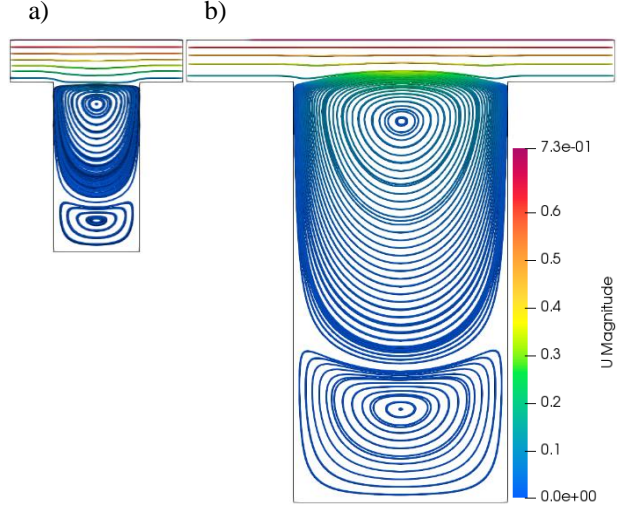

Fig. 12 Streamlines in the flow with a constant aspect ratio $w = 2.0$ a) $\frac{h_f}{d} = 0.5$ and b) $\frac{h_f}{d} = 0.2$

4.2 Pressure profile

It is now important to link the pressure distribution controlled by the speed ratio $\frac{U_{xav}}{U_x}$ to the apparent slip. Equation (8) gives the slip proportion as a function of the speed ratio $\frac{U_{xav}}{U_x}$, which can be added into Eq. (10), leading to the following equation for the pressure gradient:

$$\frac{dp}{dx} = \frac{6\mu U_x}{h_f^2}\left(2\left(\frac{\bar{\delta}_U + \frac{1}{2}}{\bar{\delta}_U + 1}\right) - 1\right) \quad (11)$$

The dimensionless pressure gradient $\frac{p_{max}-p_{min}}{dP}$ is plotted versus $\bar{\delta}_U$ for the case where the slip proportion is calculated from the average speed profile $\bar{\delta}_U(U_{xp})$ and where the slip proportion is calculated from the average inlet speed $\bar{\delta}_U(U_{xav})$ in Fig. 13. Equation (11) is plotted as a reference case. From this figure, the pressure gradient calculated from $\bar{\delta}_U(U_{xp})$ is very different from the one obtained with the modified Reynolds equation. Furthermore, the pressure gradient calculated from $\bar{\delta}_U(U_{xav})$ is always close to the pressure gradient calculated from the modified Reynolds equation. It is worth mentioning that, in this case, $\bar{\delta}_U(U_{xav})$ is always smaller than 0.1 (see Fig. 8). When the slip proportion is below 0.08, the results of the different methods are in very good correlation. This slip proportion is the maximum value reached by $\bar{\delta}_U(U_{xav})$. Higher slip values are obtained with $\bar{\delta}_U(U_{xp})$ but, for these values, the results significantly deviate from the Reynolds solution. This slip proportion calculation method is probably not relevant, as well as the one based on the shear stress that gives similar results, when estimating the pressure variation in a sliding contact.

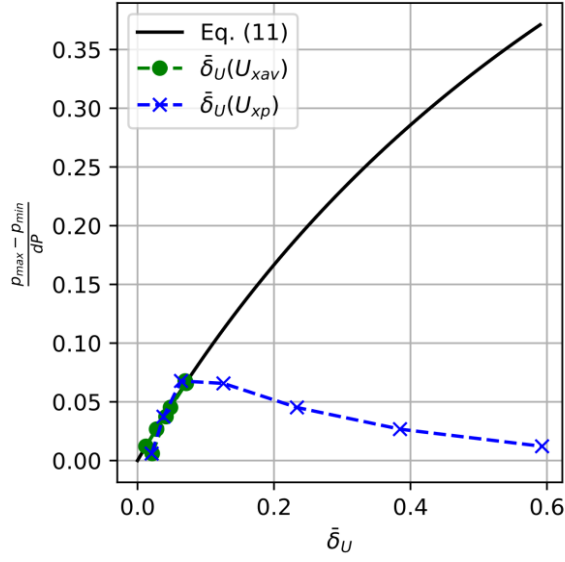

**Fig. 13 Dimensionless pressure gradient $\frac{p_{max}-p_{min}}{dP}$ with $P = \frac{6\mu U_x}{h_f^2}$ plotted versus the slip proportion. Comparison of results using $\bar{\delta}_U(U_{xp})$ and $\bar{\delta}_U(U_{xav})$ in the analytical solution**

4.3 Alternating slippery and non-slippery

Alternating slippery and non-slippery areas was carried out in a previous study, in order to produce a lift force in numerical simulations [8]. A quick numerical study was carried out for a semi-textured surface. Fig. 14 depicts the configuration of the problem. Fifty percent of the surface was textured (using three textures) and the other 50% was non-textured. The textured part was at the inlet side and the non-textured part was at the outlet side. Fig. 15 shows the pressure profiles for different size ratios and aspect ratios of the cavities. First of all, the alternation of textured and non-textured areas appears to be an efficient way to generate pressure, in accordance with Dobrica and Fillon [17]. The pressure magnitude depends on the cavity's geometrical properties. The pressure at the center of the channel is reported in Table 2, for each simulated case. The pressure gradient (Fig. 10 Dimensionless pressure gradient $\frac{p_{max}-p_{min}}{dP}$ with $P = \frac{6\mu U_x}{h_f^2}$ versus the depth ratio $\frac{h_c}{h_f}$ for different values of the size ratio $\frac{h_f}{d}$. and the slip proportion calculated from the average inlet speed and the average channel speed for a single cavity are reported in Table 2. The amount of pressure generation for the multi-cavity case is correlated with the pressure gradient value and $\bar{\delta}_U(U_{xav})$. The other slip proportion is not relevant when describing the pressure generation.

Finally, using the work of Fatu et al. [24], it is possible to analytically express the pressure in the center of the channel by using the Reynolds equation for a bearing with a slippery surface on half of its length. It is expressed as:

$$\frac{p h_f^2}{6\mu U_x l_s} = \frac{3\bar{\delta}_U}{2 + 5\bar{\delta}_U} \quad (12)$$

The results of Eq. (12) can be compared to the simulation results when $\bar{\delta}$ is replaced by $\bar{\delta}_U(U_{xav})$ and $\bar{\delta}_U(U_{xp})$ in Fig. 16. The data are those from Table 2. The lubrication results correlate very well with the simulation when $\bar{\delta}_U(U_{xav})$ is used as the slip proportion. This indicates that deep grooves can be accurately modeled by the lubrication theory, using the slip length concept based on the average inlet speed. The slip length based on the average channel speed, or based on the average wall shear stress, is not relevant here.

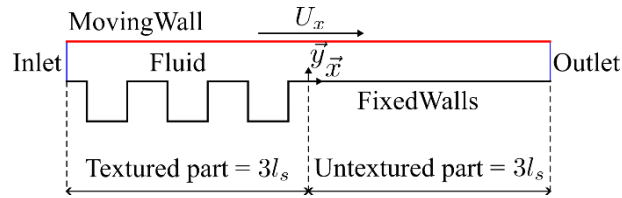

**Fig. 14 Configuration of the two-dimensional semi-textured surface in the shear flow in the numerical analysis**

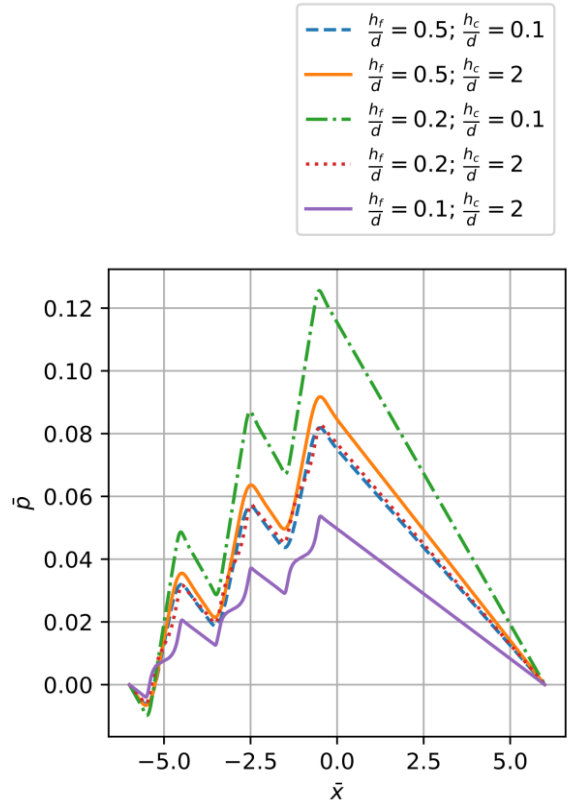

**Fig. 15** Dimensionless pressure $\bar{p} = p \dfrac{h_f^2}{6\mu U_x l_s}$ profile on the top wall versus the dimensionless coordinate $\bar{x}$ for different cavity size ratios and aspect ratios.

**Table 2** Comparison of results from the multi-cavity simulation with the mono-cavity simulation

| $\dfrac{h_f}{d}$ | $\dfrac{h_c}{d}$ | $\bar{p}\ (\bar{x}=0)$ | $\dfrac{p_{max}-p_{min}}{dP}$ | $\bar{\delta}_U(U_{xav})$ | $\bar{\delta}_U(U_{xp})$ |
|---|---|---|---|---|---|
| 0.5 | 0.1 | 0.075 | 0.058 | 0.058 | 0.049 |
| 0.5 | 2.0 | 0.084 | 0.068 | 0.070 | 0.064 |
| 0.2 | 0.1 | 0.115 | 0.098 | 0.103 | 0.074 |
| 0.2 | 2.0 | 0.077 | 0.066 | 0.071 | 0.126 |
| 0.1 | 2.0 | 0.050 | 0.045 | 0.048 | 0.233 |

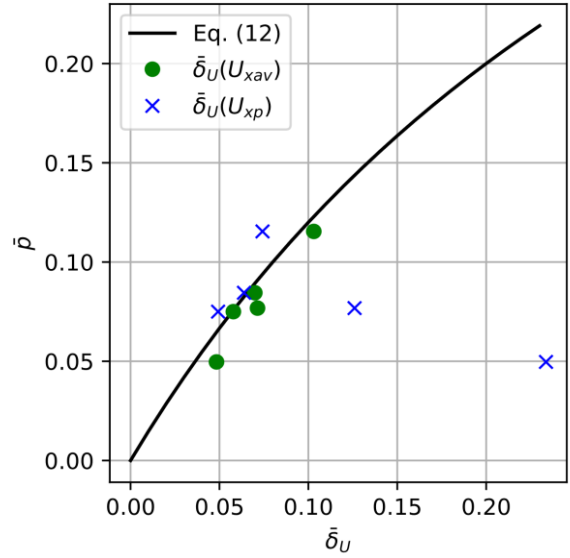

Fig. 16 Dimensionless pressure $\bar{p} = p\dfrac{h_f^2}{6\mu U_x l_s}$ in the center of the channel. Comparison of analytical solution with simulation results for the slip proportion calculated using different methods.

5 Conclusion

This paper focused on a two-dimensional study of the flow between a moving flat surface over a rectangular cavity. The density of texture and the Reynolds number, corresponding to an equivalent Couette flow with the same film height, were kept constant. The size ratio and aspect ratio are the main parameters controlling the flow in this study. Depending on the size of the cavity, a vortex can appear. Its location and size significantly affect the shear stress and the average flow. The effect of the cavity can be analyzed using the notion of apparent slip length. The slip phenomenon can be seen as a macroscopic representation of a phenomenon which is not slippery at the local level. Three methods to calculate the slip length based on the average wall shear stress, the average channel speed or the average inlet speed, are compared. Even if the two former methods give identical results, the latter appears to be more consistent when compared to the lubrication theory.

The pressure profile on the moving wall was also analyzed but it was difficult to confirm a lift force for a single cavity bonded by a periodic condition. Thus, the pressure gradient, calculated from the minimal to maximal relative pressures and from the cavity width, was analyzed. This pressure gradient is consistent with the lubrication theory when it is presented as a function of slip proportion, based on the average inlet speed of the fluid $\bar{\delta}_U(U_{xav})$.

An example of the application of the present analysis is given for a shear flow between a smooth surface and a semi-textured surface. It is shown that a lift force is generated and the generated pressure is correlated with the single cavity results. The description of this case with the lubrication is possible when the slip proportion is based on the average inlet speed of the fluid $\bar{\delta}_U(U_{xav})$. Calculating the slip length with the shear stress is not relevant for this problem. Future work should also be carried out on multiphasic flows in a Cassie-Baxter wetting state (with air entrapped in the cavities) and three-dimensional flows for a better analysis of the notion of slip length. In addition, some experiments on textured hydrophobic surfaces in shear and pressure flows will be performed to consolidate the numerical results.

# 6 Acknowledgments

This work pertains to the French government program 'France 2030' (LABEX INTERACTIFS, reference ANR-11-LABX-0017-01, and EUR INTREE, reference ANR-18-EURE-0010). The authors would like to acknowledge Région Nouvelle Aquitaine for funding this research via the TECAP Project.

**8 Table and Figure Captions**

Table 1 Comparison of the slip length calculation methods

Table 2 Comparison of results from the multi-cavity simulation with the mono-cavity simulation

Figure 17 Configuration of the two-dimensional cavity for the numerical analysis.

Figure 18 a) Speed profile of the fluid and b) slip length definition

Figure 19 Comparison of the non-dimensional pressure profile with the one obtained by Arghir et al. [18] versus the non-dimensional $\bar{x}$ axis. The non-dimensional parameters are $\frac{d}{l_s} = 0.5$, $\frac{h_c}{h_f} = 2.0$ and $Re_{h_0} = \frac{\rho U_x h_0}{\mu} = 100$ with $h_0 = h_f + \frac{h_c}{2}$. The continuous line represents the reference and the dotted line represents the comparison from the present model.

Figure 20 Non-dimensional local wall shear stress $\bar{\tau}_l = \frac{\tau_l h_f}{\mu U_x}$ on the moving wall for different size ratios of the cavity as a function of the non-dimensional coordinate $\bar{x}$ with a constant aspect ratio $w = 2$.

Figure 21 Non-dimensional averaged wall shear stress on the moving wall $\bar{\tau}$ as a function of the aspect ratio $\frac{h_c}{d}$ for different values of the size ratio $\frac{h_f}{d}$.

Figure 22 Non-dimensional wall shear stress $\bar{\tau}$ as a function of the depth ratio $\frac{h_c}{h_f}$ for different values of the size ratio $\frac{h_f}{d}$.

Figure 23 Streamlines and speed (m/s) in a cavity for a constant size ratio $\frac{h_f}{d} = 0.20$ with a) $w = \frac{h_c}{d} = 0.10$, b) $w = 0.16$ and c) $w = 1.60$ and a constant size ratio $\frac{h_f}{d} = 0.5$ where d) $w = 0.25$, e) $w = 0.40$, and f) $w = 1.50$.

Figure 24 Comparison of the slip proportion calculated from wall shear stress $\bar{\delta}_\tau$, with that calculated from the average speed of the fluid $\bar{\delta}_U(U_{xav})$ and the one calculated from the average speed profile in the lubricant film fluid $\bar{\delta}_U(U_{xp})$ versus the size ratio with a constant aspect ratio $w = 2$.

Figure 25 Dimensionless pressure $\bar{p} = p \frac{h_f^2}{6\mu U_x l_s}$ on the moving wall for different size ratios of the cavity versus the non-dimensional coordinate $\bar{x}$ with a constant aspect ratio $w = 2$.

Figure 26 Dimensionless pressure gradient $\frac{p_{max} - p_{min}}{dP}$ with $P = \frac{6\mu U_x}{h_f^2}$ versus the depth ratio $\frac{h_c}{h_f}$ for different values of the size ratio $\frac{h_f}{d}$.

Figure 27 Dimensionless pressure gradient $\frac{p_{max} - p_{min}}{dP}$ with $P = \frac{6\mu U_x}{h_f^2}$ versus the dimensionless inlet fluid speed for different size ratios of the cavity.

Figure 28 Streamlines in the flow with a constant aspect ratio $w = 2.0$ a) $\frac{h_f}{d} = 0.5$ and b) $\frac{h_f}{d} = 0.2$.

Figure 29 Dimensionless pressure gradient $\frac{p_{max} - p_{min}}{dP}$ with $P = \frac{6\mu U_x}{h_f^2}$ plotted versus the slip proportion. Comparison of the results with $\bar{\delta}_U(U_{xp})$ and $\bar{\delta}_U(U_{xav})$ to the analytical solution.

Figure 30 Configuration of the two-dimensional semi-textured surface in the shear flow for the numerical analysis.

Figure 31 Dimensionless pressure $\bar{p} = p \frac{h_f^2}{6\mu U_x l_s}$ profile on the top wall versus the dimensionless coordinate $\bar{x}$ for different cavity size ratios and aspect ratios.

Figure 32 Dimensionless pressure $\bar{p} = p \frac{h_f^2}{6\mu U_x l_s}$ in the center of the channel. Comparison of analytical solution with simulation results for slip proportion calculated with different methods.